\documentclass[letterpaper, 10 pt, conference]{ieeeconf}
\usepackage{amsmath,amsfonts}
\usepackage{algorithmic}
\usepackage{array}
\usepackage{textcomp}
\usepackage{stfloats}
\usepackage{url}
\usepackage{verbatim}
\usepackage{graphicx}
\usepackage{bbold}
\usepackage{balance}
\usepackage{cite}
\usepackage[font=footnotesize]{caption} 
\usepackage{subfigure}
\usepackage{array}
\usepackage{makecell}
\usepackage{xcolor}
\newtheorem{theorem}{Theorem}
\newtheorem{proposition}{Proposition}

\newcommand{\lba}{\left[ \begin{array}}
\newcommand{\ear}{\end{array} \right]}

\IEEEoverridecommandlockouts                        
\title{Implications of Grid-Forming Inverter Parameters on Disturbance Localization and Controllability 
\thanks{* Corresponding author. This work was supported by National Science Foundation (NSF) under Grant ECCS-2527653.}
}

\author{
    \authorblockN{
        Matt Baughman,\;
        Marena Trujillo,\;
        Bri-Mathias Hodge,\;
        Emily Jensen*
    }
    \authorblockA{
        Department of Electrical, Computer \& Energy Engineering\\
        University of Colorado Boulder \\
        Boulder, Colorado, United States\\
        \{matt.baughman, marena.trujillo, brimathias.hodge, ejensen\}@colorado.edu
    }
}

\begin{document}

\maketitle
\begin{abstract}
The shift from traditional synchronous generator (SG) based power generation to generation driven by power electronic devices introduces new dynamic phenomena and  considerations for the control of large-scale power systems. In this paper, two aspects of all-inverter power systems are investigated: greater localization of system disturbance response and increased system controllability.  The prevalence of both of these aspects are shown to be related to the lower effective inertia of inverters and have implications for future wide-area control system design. Greater disturbance localization implies the need for feedback measurement placement close to generator nodes to properly reject disturbances in the system, while increased system controllability implies that wide-area control systems should preferentially actuate inverters to most efficiently control the system. This investigation utilizes reduced-order linear time-invariant models of both SGs and inverters that are shown to capture the frequency dynamics of interest in both all-SG and all-inverter systems, allowing for the efficient use of both frequency and time domain analysis methods. 
\end{abstract}
\section{Introduction}

The share of power-electronic generation has grown rapidly in the 21st-century as solar PV, wind, and battery storage, all of which require an inverter, have become cost-competitive with conventional generation \cite{Lazard2024}. Inverter-based resources (IBRs) are generally classified as Grid-Following (GFL) or Grid-Forming (GFM) \cite{linResearchRoadmapGridForming}. GFLs regulate AC current and rely on a phase-locked loop (PLL) to track the grid frequency \cite{Li2022}, meaning another device must “form’’ the grid. Synchronous generators (SGs) naturally provide this function by establishing their own voltage and frequency \cite{SauerStabilityTextbook}. Today, GFLs operate in mixed SG–IBR systems where SGs supply the grid-forming capability \cite{Li2022}. GFMs, in contrast, establish voltage and frequency through closed-loop control and are therefore essential for all-IBR power systems \cite{linResearchRoadmapGridForming}. Among various GFM implementations, this paper focuses on droop-controlled GFMs that use $P$–$\omega$ droop to maintain synchronization \cite{Darco2014}.

While the dynamics of systems of SGs have been widely analyzed \cite{SauerStabilityTextbook}, legacy methods may not be applicable to IBR-dominated power systems characterized by fast dynamics and low inertia values \cite{Hatziargyriou2021}.  The fast dynamics from IBRs and the increasing deployment of new measurement systems in large-scale power systems available for feedback control has made the implementation of wide-area control systems (WACs) both necessary and feasible \cite{Chakrabortty2013}. To inform wide-area control design decisions, an interpretable connection between IBR design parameters and IBR-dominated system behavior and controllability is necessary.

System controllability refers to how much energy is needed to control the system along a state trajectory and can be quantified by the eigenvalues of the controllability gramian \cite{Lindmark2018}. Many studies have evaluated the controllability of systems of SGs \cite{pasqualetti2014, li2015critical, Li2016}, and more recently, there have been efforts to assess the controllability of systems with some amount of IBRs. For instance, in \cite{Roy2021}, the controllability of a 100\% IBR microgrid is assessed to determine how controller flexibility may be utilized to support small-signal  stability. In \cite{zhan2021}, the modal controllability matrix of a power system with wind turbines is used for the optimal aggregation of an impedance network model. Analytical expressions for performance metrics related to frequency stability are presented in \cite{Paganini2020} under the assumption of proportional machine parameters. In \cite{ducoin2023swing}, GFM and GFL frequency dynamics are modeled using modified swing equations to explain their differing inertial responses. This paper compliments these recent works by providing a connection between GFM  parameters and system disturbance response and controllability, allowing for an interpretable connection between GFM design decisions and system behavior.
This paper presents a novel assessment of the impact of effective GFM inertia on disturbance localization and system controllability. Increased disturbance localization under high GFM penetration scenarios is demonstrated, motivating the need for strategic feedback measurement locations in a wide-area control policy. Additionally, evidence of the increased controllability of GFM-dominated systems is presented, indicating the importance of actuating GFMs for the efficient control of future system dynamics. 
%

%

\subsection{Mathematical Notation}
\color{black}
    $\mathbb{R}$ denotes the set of real numbers and $\mathbb{R}^{n \times m}$ denotes the set of real-valued matrices of dimension $n \times m$. $M^{\top} \in \mathbb{R}^{n \times m}$ denotes the transpose of a matrix $M \in \mathbb{R}^{m \times n}$. We write $P \succ 0$ to denote that $P$ is a positive definite matrix.
    $I_n$ denotes the $n \times n$ identity matrix, $\mathbb{0}_{n \times m}$ denotes the matrix of all zeros of dimension $n \times m$, and  $\mathbb{1}_{n \times m}$ denotes a matrix of all ones of dimension $n \times m$. We denote $\mathbb{1}_n = \mathbb{1}_{n \times 1},$ and we omit subscripts when dimensions are clear from context. 
    
    $j$ denotes the imaginary unit $j = \sqrt{-1}$. If $z = \alpha + j \beta$ is a complex number, then $z^* = \alpha - j \beta$ denotes its complex conjugate and $|z|$ denotes its magnitude.
\color{black}

\section{Reduced Order Model} \label{sec:model_info}
This paper provides interpretable analytical relationships between grid-forming inverter (GFM) parameters and system behavior.  To achieve this, we leverage the reduced-order linear time invariant (LTI) models of GFM frequency dynamics proposed in \cite{trujilloAnalyticalModelsFrequency2026} and the second-order synchronous generator (SG) dynamics of \cite{sajadiSynchronizationElectricPower2022}.  This section offers a physical interpretation of these model parameters that will motivate the susequent analysis. 
\subsection{Model Summary}
The network representation is based on the Kron-reduced DC power flow equation, where only generator nodes are retained \cite{trujilloAnalyticalModelsFrequency2026}. The expression for the change in power injection at the retained generator nodes is given by
\begin{equation} \label{eq:dPG}
     \Delta P = \boldsymbol{B}_{r}\Delta\delta_G+\Delta P_d,
\end{equation}
\color{black}{where $\Delta \delta_G$ denotes a column vector of deviations from nominal voltage angle at retained generator buses, $\Delta P_d$ is the column vector of active power deviations at generator nodes due to a disturbance, and $\mathbf{B_r}$ is the reduced network susceptance matrix.}\color{black} 

\begin{figure}[h!]
    \centering
    \includegraphics[width=.45\linewidth,angle = -90]{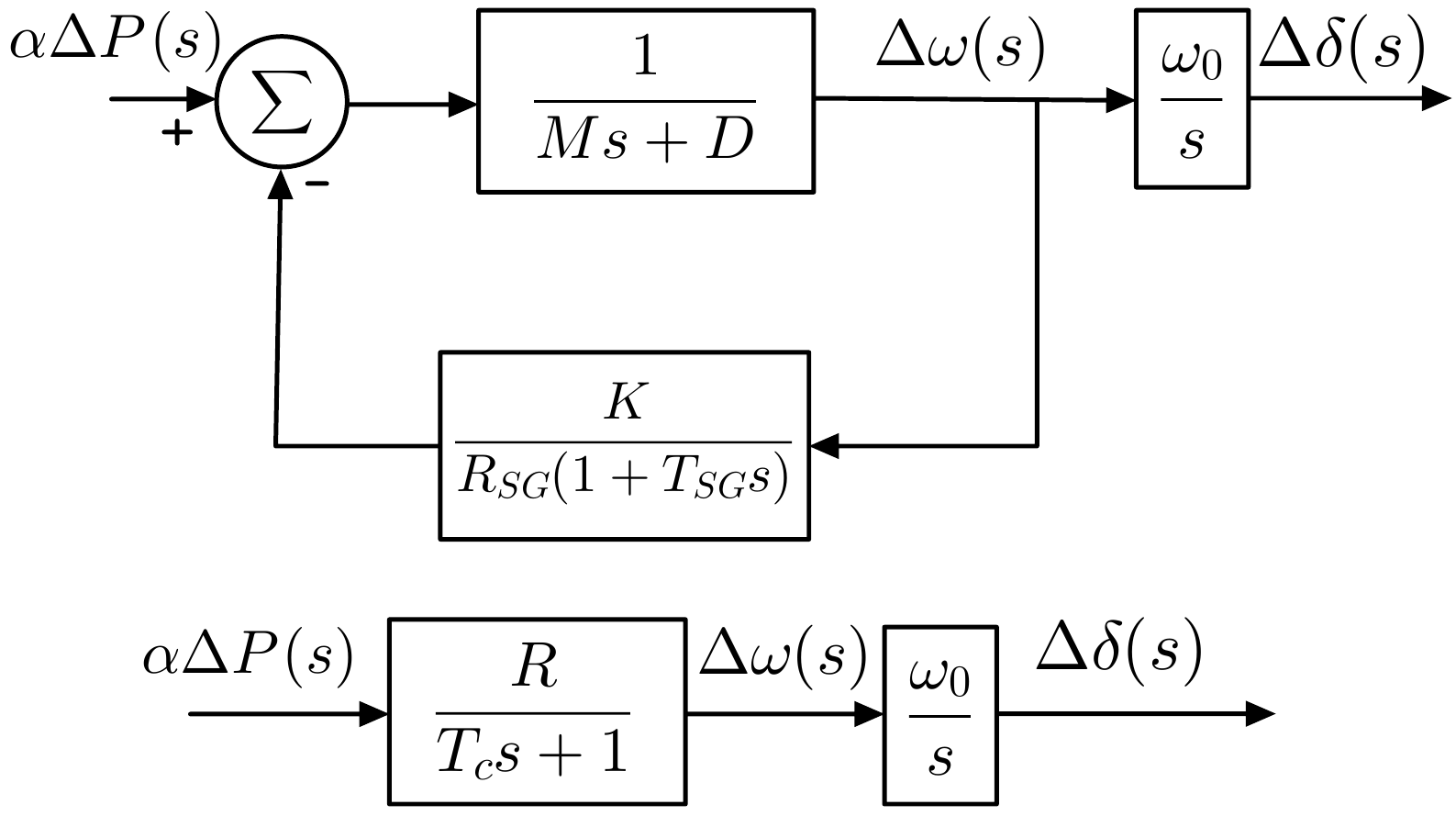}
    \caption{Block diagram for the SG model (top)\cite{sajadiSynchronizationElectricPower2022}, and droop-control GFM (bottom) \cite{trujilloAnalyticalModelsFrequency2026}.}
    \label{fig:SG_block}
    \vspace{-5mm}
\end{figure}

The SG model used is depicted in block diagram form in Figure \ref{fig:SG_block} and its transfer function is given by
\begin{equation}\label{eq:SGtf}
    \frac{\Delta \omega(s)}{\alpha \Delta P(s)} = \frac{sT_{SG} + 1}{s^2T_{SG}M\!+\!s(DT_{SG}+M) + (D+\frac{K}{R_{SG}})}.
\end{equation} 
$M$ and $D$ refer to the inertia and damping coefficients of the generator's shaft, respectively. \textcolor{black}{The momentum, $M$, is twice the inertia constant, $H$ of the machine: $M = 2H$.} The governor is connected in feedback with an input of frequency. $K$ represents the governor and turbine response gain, $R_{SG}$ is the droop coefficient, and $T_{SG}$ is the governor and turbine response time constant. 

The model of a droop GFM is shown in block diagram form in Figure \ref{fig:SG_block}.  $R$ represents the GFM droop coefficient and $T_C$ represents the time constant of the GFM frequency response \cite{trujilloAnalyticalModelsFrequency2026}. \textcolor{black}{All parameters, in both the SG and GFM models, are per-unit values.} The parameter $\alpha$ is the ratio of the system base power to the device rated capacity\cite{trujilloAnalyticalModelsFrequency2026} and $\omega_0$ is the nominal generator frequency. The GFM model utilized is a first-order transfer function with the following relationship between change in GFM power and change in GFM frequency:
\begin{equation}
    \frac{\Delta \omega (s)}{\alpha \Delta P(s)} = \frac{1}{\frac{T_c}{R}s+\frac{1}{R}}
\end{equation}

%

In the SG model, the transfer function representing the shaft dynamics maps a change in electrical power to a change in frequency, while in the GFM the outer-loop droop controller is responsible for this mapping. In both the SG and GFM models, this transfer function is first order, and this parallel can be used to understand GFM dynamics in a more traditional framework. In this framework, the effective inertia of the droop GFM is equal to $M_{eff} = \frac{T_c}{R}$ and the effective damping defined as $D_{eff} = \frac{1}{R}$. Under the assumption that all GFMs have identical effective inertia, damping and rated capacity, a system of $n$ GFM nodes and $m$ load nodes can be written as a system of ODEs with the following state vector:
\begin{equation}
\small
    x = \begin{bmatrix}
        (\Delta\delta_{1}-\Delta\delta_{n}) 
        \dots 
        (\Delta\delta_{n-1}-\Delta\delta_{n}) &&&&
        \dot{\Delta\omega_1} &
        \dots &
        \dot{\Delta\omega_n}
    \end{bmatrix}^T
\end{equation}
\begin{subequations}
\begin{align}
        &\dot{x}= A x + B \Delta P_d,\\
        & A =     \begin{bmatrix}
        \mathbb{0} & \,\,\omega_0\begin{bmatrix} I_{n-1} & \mathbb{1}_{n-1}\end{bmatrix}\\
\frac{-\alpha }{M_{eff}}\boldsymbol{\tilde{B_{r}}} & \frac{-D_{eff}}{M_{eff}}I_n
    \end{bmatrix},  B = \begin{bmatrix}
        \mathbb{0}\\
       \frac{\alpha}{M_{eff}} I_{m}
    \end{bmatrix},
\end{align}  \label{eq:state_space_GFM}\end{subequations}
where $\tilde{\boldsymbol{B_r}} \in \mathbb{R}^{n \times (n-1)}$ refers to $\boldsymbol{B_{r}}$ \textcolor{black}{with its last column, which corresponds to the reference bus, removed.} 

We assume a time constant of $T_C = 0.0318 $s and a droop coefficient of $R = 0.05$ (or 5\% droop), which yields an effective inertia of $M_{eff} = 0.636$ and effective damping of $D_{eff} = 20$.  Compared to standard SG inertia and damping values which range from 2 to 9 and 1 to 2, respectively\cite{SauerStabilityTextbook}, the GFM effective inertia is roughly an order of magnitude smaller than usual SG inertia values while the effective damping is roughly an order of magnitude greater; however, parameters can be tuned to modify the GFM's response due to the digital nature of GFM controls \cite{QiuOscillationSuppression}. An analytical interpretation of how these GFM parameters impact system behavior is discussed in the next section.

\section{Analysis of Disturbance Localization in All-GFM Systems} \label{sec:analytical_dist} 
To create an interpretable relationship between system behavior and GFM parameters, a three-GFM system model provides an illustrative example. \textcolor{black}{As is common for analytical analysis of power systems \cite{GrossCompensatingNetworkDynamics,siegelmann2025stability}, this model assumes uniform line admittance coupling between generators, such that the reduced susceptance matrix can be written as }

\begin{equation}\label{eq:B}
\setlength{\arraycolsep}{5pt}
    \mathbf{B_r} = b 
    \begin{bmatrix}
        2&-1&-1\\
        -1&2&-1\\
        -1&-1&2
    \end{bmatrix} .
\end{equation}
All GFMs share the same effective inertia and damping which, for readability, are written as $M$ and $D$ respectively based on the physical interpretation of these parameters discussed in Section \ref{sec:model_info}.  Here, we derive and compare the transfer functions of a generator's response
to a disturbance occurring 1) at the generator's bus, and 2) at a different generator's bus. 
The transfer functions are used to derive the poles and zeros of the system for both the local and nonlocal disturbance response of the GFMs in the system. The location of the poles and zeros can then be used to understand how GFM parameters relate to the localization of disturbances. The derived transfer functions will also be used to form analytical expressions for the \textcolor{black}{$\mathcal{H}_2$-norm} of the GFM's local and nonlocal response, a common metric used to measure responsiveness in power systems \cite{Tegling2015}, to further show how effective GFM inertia impacts responses to local and nonlocal disturbances.

\color{black}Due to symmetry of the network, \color{black} the derived expressions relating the changes in net generator power at each generator node to changes in generator frequency form a $3\times3$ symmetric, circulant transfer function matrix \color{black} with only two distinct transfer function entries\color{black} -- the diagonal terms:
\begin{equation}
\label{p2w_tf_mat}
     T_{\rm d}(s): =   \frac{\Delta\omega_l(s)}{\Delta P_{d,l}(s)} = \frac{\alpha \,{\left(M\!s^2\!+D\,s+\!b\,\alpha \,\omega_0 \right)}}{{\left(D+\!M\,s\right)}\,{\left(Ms^2\!+D\,s+\!3\,b\,\alpha \,\omega_0 \right)}}
 \end{equation}
 for $\ell = 1,2,3$ and the off-diagonal terms:
 \begin{equation}
      T_{\rm od} (s): =\! \frac{\Delta\omega_l(s)}{\Delta P_{d,k}(s)}\!=\!\frac{b\,\alpha^2 \,\omega_0 }{{\left(D\!+\!Ms\right)}\,{\left(Ms^2\!+D\,s+\!3b\alpha \,\omega_0 \right)}},
\end{equation}
for $l \neq k, ~ l,k \in \{1,2,3\}$.

\color{black}
\subsection{Pole-Zero Analysis} 
\begin{figure}
    \centering
    \includegraphics[width=\linewidth]{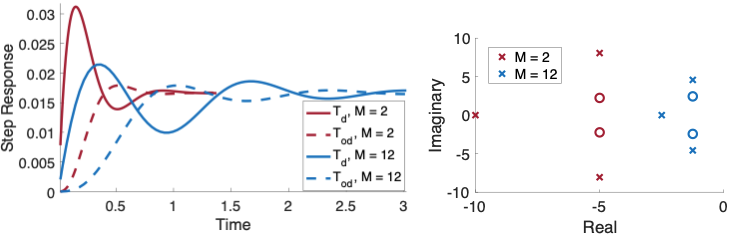}
    \caption{(Left) Step response of $T_{\rm d}$ and $T_{\rm od}$ for values of $M = 2, 12$. (Right) Pole-zero plot for $T_{\rm d}$ for values of $M = 2, 12.$}
    \label{fig:pole_zero}
    \vspace{-4mm}
\end{figure}

\color{black}
Both $T_{\rm d}(s)$ and $T_{\rm od}(s)$ have three poles:
\begin{equation}
p,p^* = -\frac{D\pm j \sqrt{-{D}^2 +12\,b\,M\,\alpha \,\omega_0 }}{2\,M},~ \tilde{p} = 
 \frac{-D}{M}   
\label{eq:pole_exp}
\end{equation} 
while only $T_{\rm d}(s) $ has two zeros: \begin{equation}
z, z^* = -\frac{D\pm j\sqrt{-{D}^2 +4\,b\,M\,\alpha \,\omega_0 }}{2\,M},
\label{eq:zero_exp}
\end{equation}
which are depicted in Figure~\ref{fig:pole_zero}.
For physically relevant parameter values of interest, $(-{D}^2 +4\,b\,M\,\alpha \omega_0)$ will be a negative number, so that the magnitude and real component of the zeros are given by 
\begin{equation}
    |z| = \sqrt{\frac{b \alpha \omega_0}{M}}, ~~
    {\rm Re}(z) = \frac{-D}{2M}, \end{equation}
and the magnitude and real component of the complex-valued poles are given by 
\begin{equation}
    |p| = \sqrt{\frac{3 b \alpha \omega_0}{M}} = \sqrt{3} |z|, ~~ {\rm Re}(p) = \frac{-D}{2M} = {\rm Re}(z).
\end{equation}
Note that there will never be a pole-zero cancellation. The additional left half plane (LHP) zeros in $T_{\rm d}$ relative to $T_{\rm od}$ will lead to a ``more reactive" local response (see Figure~\ref{fig:pole_zero}). 

\color{black}
To view how the responses of diagonal and off diagonal entries vary with $M$, we examine the step responses of $T_{\rm od}$ and $T_{\rm d}$, given by the inverse Laplace transform of: 
\begin{equation} \begin{aligned}
    \frac{T_{\rm d}(s)}{s} & = \frac{g_{\rm d}}{s-p} + \frac{g_{\rm d}^*}{s-p^*} + \frac{h_{\rm d}}{s-\tilde{p}} + \frac{k_{\rm d}}{s}, \\
    \frac{T_{\rm od}(s)}{s} & = \frac{g_{\rm od}}{s-p} + \frac{g_{\rm od}^*}{s-p^*} + \frac{h_{\rm od}}{s-\tilde{p}} + \frac{k_{\rm od}}{s}.
\end{aligned} \end{equation}
\color{black}{As the GFM effective inertia decreases, the complex poles $\{p, p^*\}$ move further left in the complex plane and further from the real axis, leading to faster decay and greater oscillations.} The corresponding residues of the diagonal response relative to those of the the off-diagonal response, $\tfrac{g_{\rm d}}{g_{\rm od}}$, vary with the zero locations and grow like $\tfrac{1}{M}$ as $M$ decreases with decreasing GFM effective inertia. The real-valued pole moves farther left leading to more rapid decay. \color{black}

This analysis supports the observed step responses, which demonstrate higher overshoot and faster decay as $M$ decreases (see Figure~\ref{fig:pole_zero}). We interpret this as follows. A given generator will be more reactive to an input at that generator's bus (emulating a disturbance close to the generator in a more realistic topology) than to an input at a different generator bus (emulating a disturbance far from the generator in a more realistic topology). This reactivity increases as $M$ decreases.

\indent 
%
The preceding analysis suggests that the generator frequencies of higher inertia power systems, like those dominated by SGs, will generally move together, while low inertia power systems will see greater divergence in frequency in buses closer to the location of the disturbance.
\subsection{$\mathcal{H}_{2}$--norm Analysis}
The $\mathcal{H}_{2}$-norm, defined as the root-mean square of a system's impulse response,
\begin{equation}
    \left \lVert {H(s)}\right \rVert_{\mathcal{H}_2}^2=\frac{1}{2\pi}\int_{-\infty}^{\infty} \left \lVert {H(j\omega)}\right \rVert^2d\omega,
    \label{eq:H2_norm}
\end{equation}
is calculated for each of the transfer functions $T_{\rm d}$ and $T_{\rm od}$. A proof of this result is provided in the Appendix. 

\begin{proposition} \label{prop:H2}
The $\mathcal{H}_2$-norm of the diagonal and off-diagonal transfer functions, $T_{\rm d}$ and $T_{\rm od}$, for the simple three-GFM network are given by:
\begin{subequations} 
	\begin{align}
    & \|T_{\rm d}\|_{\mathcal{H}_2}^2 = \frac{\alpha^2 {\left(6{D}^2 +5M\alpha b\omega_0 \right)}}{6DM{\left(2{D}^2 +3M\alpha b\omega_0 \right)}}
    \label{eq:localH2}\\
    & \|T_{\rm od}\|_{\mathcal{H}_2}^2=\frac{\alpha^3 b\omega_0 }{6{D}^3 +9M\alpha b\omega_0 D}. \label{eq:nonlocalH2}
	\end{align} 
\end{subequations}
\vspace{-10pt}
\end{proposition}

\textcolor{black}{From Proposition~\ref{prop:H2}, we observe that the size of $T_{\rm d}$ relative to the size of $T_{\rm od}$, $\frac{\|T_{\rm d}\|_{\mathcal{H}_2}}{\|T_{\rm od}\|_{\mathcal{H}_2}}$, increases monotonically as the effective GFM inertia decreases. In line with our pole-zero interpretation, we interpret this increase in $\frac{\|T_{\rm d}\|_{\mathcal{H}_2}}{\|T_{\rm od}\|_{\mathcal{H}_2}}$ as greater localization in the disturbance response. We conjecture that this disturbance response localization should influence the selection of measurement locations to be used for feedback under a WAC policy. }

\section{System Controllability in All-GFM Systems} 

\label{sec:analytical_controllability}
We now analyze how the eigenvalues of the system's controllability gramian move with respect to system parameters. Here we assume each GFM has adequate headroom for frequency response and that a WAC system is able to change the power setpoint of each GFM in the network directly. The controllability gramian, $W_C$, is the positive definite solution to the Lyapunov equation
\begin{equation} \label{eq:ctrb_gram}
   AW_C + W_CA^T = -BB^T,
\end{equation}
where $A$ and $B$ are defined in equation \eqref{eq:state_space_GFM}. The eigenvalues and eigenvectors of the controllability Gramian define an ellipsoid that represents the set of states reachable with unit-energy input.  The ellipsoid has semiaxis lengths equal to the square root of the eigenvalues and semiaxis directions given by the corresponding eigenvectors \cite{Lindmark2018}.  This implies that larger eigenvalues of the controllability Gramian correspond to directions of a system that are easier to control \cite{Kalman1962}. For a system of GFMs, the controllability Gramian eigenvalues and their multiplicity are given by the following theorem: 

\begin{theorem} \label{main_thm} Given a network of $n$ grid-forming inverters with identical effective inertia, $M$, identical effective damping $D$, and identical line susceptance coupling each inverter to every other inverter in the network, $b$, the eigenvalues of the controllability Gramian and their multiplicity are: 
\begin{itemize}
    \item $\frac{\omega_0}{2bD}$ with multiplicity 1 associated with controlling relative GFM angles
    \item $\frac{\omega_0}{2nbD}$ with multiplicity $n-2$ associated with controlling relative GFM angles
    \item $\frac{1}{2MD}$ with multiplicity $n$ associated with controlling GFM frequencies
\end{itemize}
\end{theorem}

A proof is given in the Appendix.  This theorem implies that lower effective GFM inertia \emph{increases the controllability} of GFM frequencies from the perspective of a wide-area control system \textcolor{black}{and does not impact controllability of relative angle differences}. This direct connection between effective inertia and controllability will influence future work developing WAC design methods, implying that actuating GFMs will be a more effective way to control system frequencies.

\section{Computational Analysis of Larger Systems} \label{sec:larger_sys}
This section supports the conclusions about increased localization and controllability from  Sections \ref{sec:analytical_dist} and \ref{sec:analytical_controllability} using larger, more realistic power system models -- specifically the IEEE 39-bus and IEEE 118-bus models. \textcolor{black}{These models feature heterogeneous generator capacities and line susceptances.} The disturbance localization analysis is performed by comparing time-domain frequency trajectories at different generators and by comparing the $\mathcal{H}_2$-norm from each load bus input to each generator frequency output.
 The controllability analysis is performed by numerically finding the controllability gramian eigenvalues for varying effective inertia. \textcolor{black}{All analysis is done using MATLAB.}
\subsection{Disturbance Localization Time Domain Analysis}
\begin{figure}[h!]
    \centering
    \includegraphics[width=0.9\linewidth]{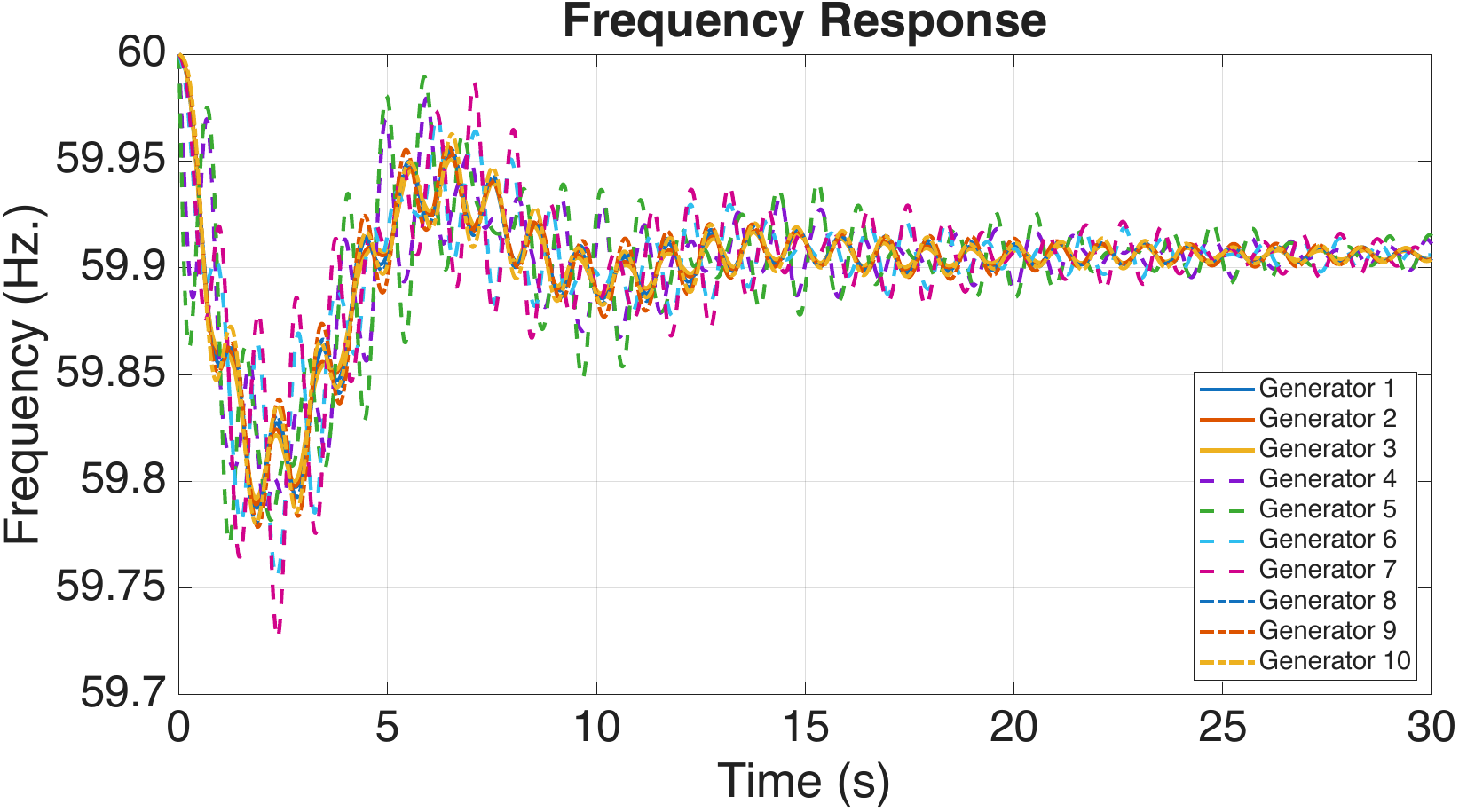}
    \caption{IEEE 39-Bus All-SG Frequency Response for Step at Bus 20}
    \label{fig:SG_freq_resp_39bus}
\end{figure}

\begin{figure}[h!]
    \centering
    \includegraphics[width=0.9\linewidth]{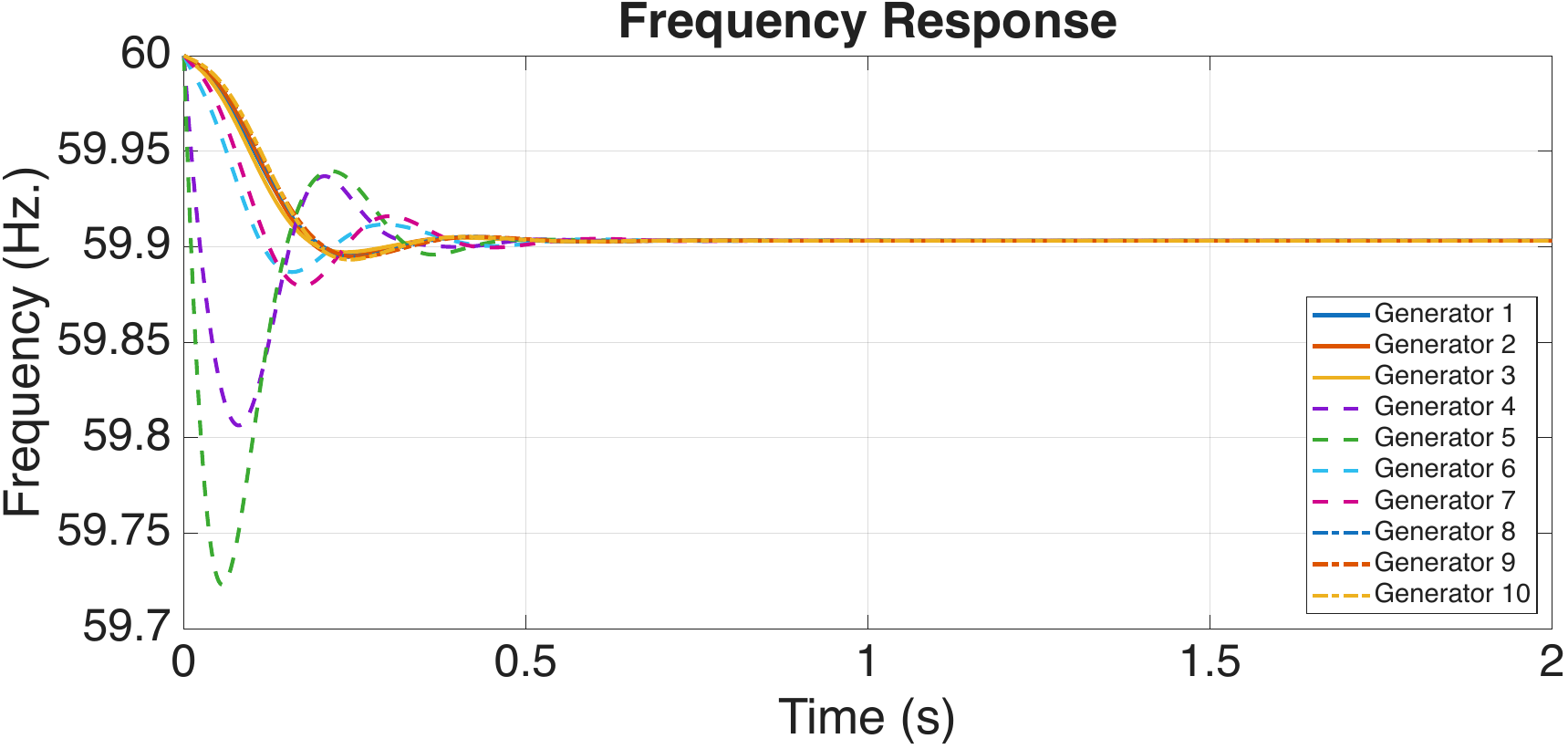}
    \caption{IEEE 39-Bus All-GFM Frequency Response for Step at Bus 20}
    \label{fig:GFM_freq_resp_39bus}
\end{figure}

A load step is applied to Bus 20 in the IEEE-39 bus system, and the resulting frequency dynamics of each of the generators is shown in Figures \ref{fig:SG_freq_resp_39bus} and \ref{fig:GFM_freq_resp_39bus}. The dynamics of the all-SG system follows the expected second-order system response of a traditional power system, with generators closer to the disturbance (generators 4 and 5) displaying greater oscillations on top of the overall system's frequency response.  However, the response of the all-GFM system shows a greater heterogeneity in frequency response, with generators closer to the disturbance bus transiently diverging from the rest of the system. This indicates that all-GFM systems may experience greater localization in their load step response as compared to all-SG systems. 

\subsection{Disturbance Localization  Analysis}
\begin{figure}[h!]
    \centering
    \subfigure[] {\includegraphics[width=0.47\linewidth]{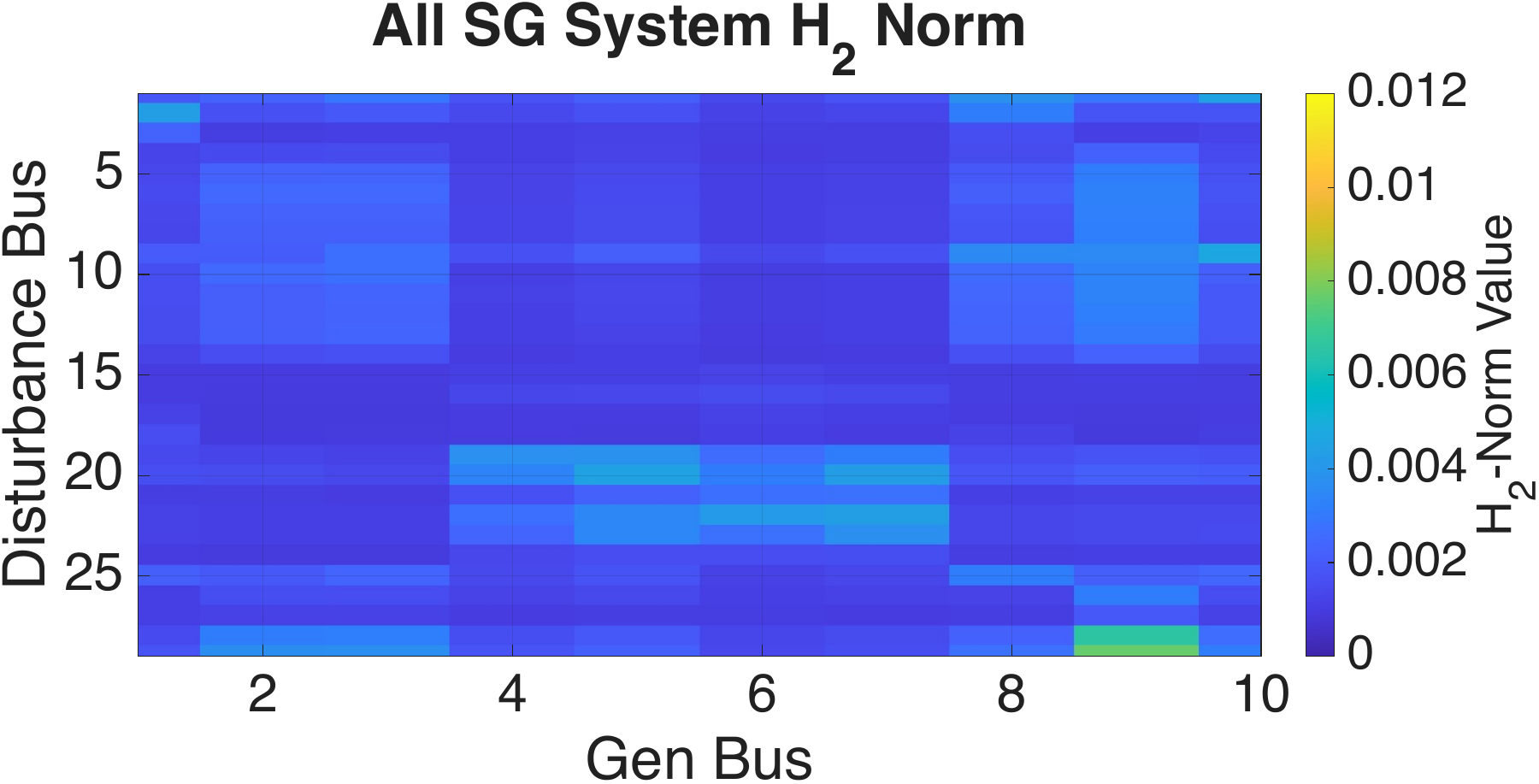}}
     \subfigure[] {\includegraphics[width=0.47\linewidth]{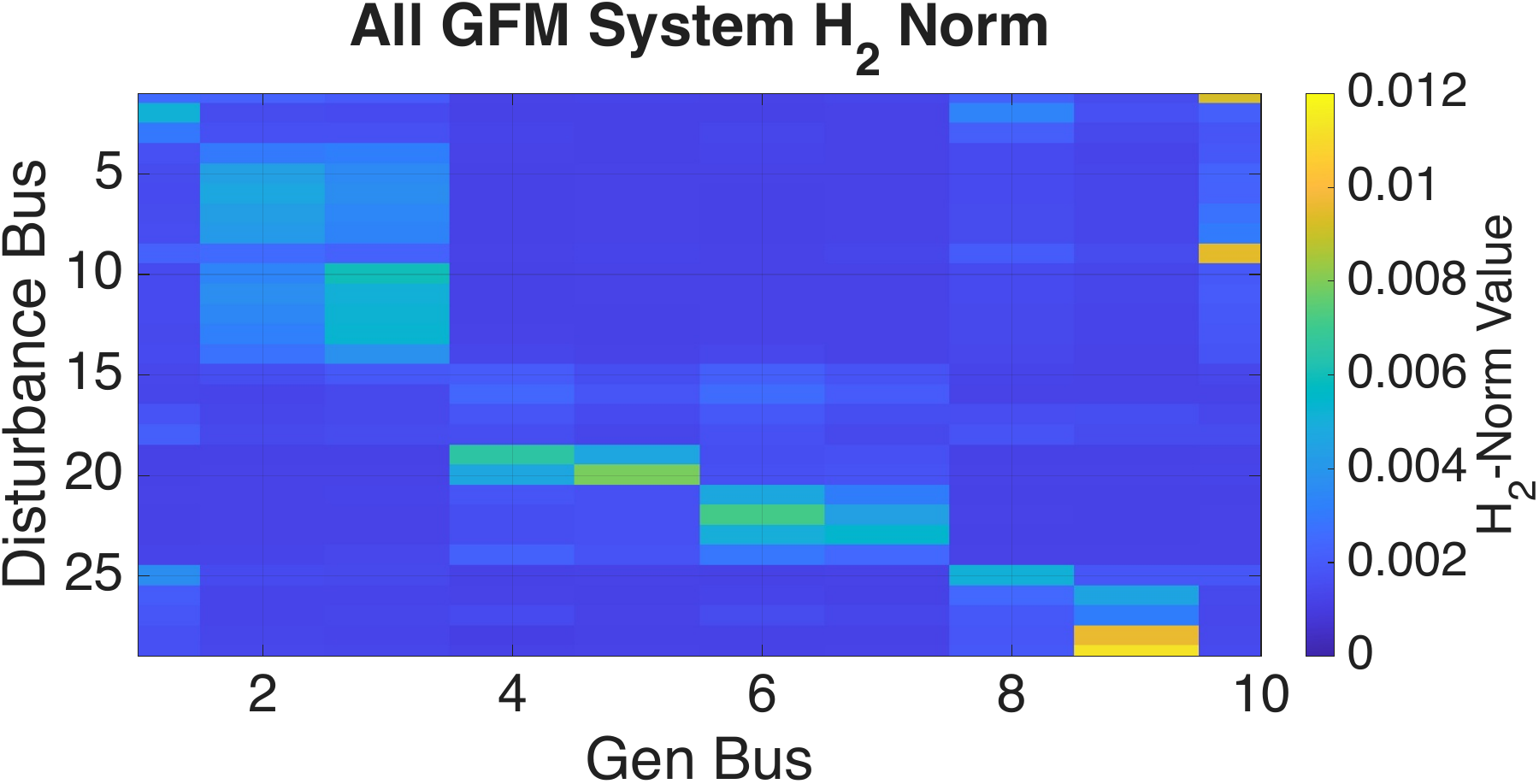}}
    \caption{Heat map plot depicting the $\mathcal{H}_2$-norms for the IEEE-39 bus system for (a) All-SG System and (b) All-GFM System.}
    \label{fig:h2_norm_map} \vspace{-4mm}
\end{figure}

\begin{figure}[h!]
    \centering
     {\includegraphics[width=0.8\linewidth]{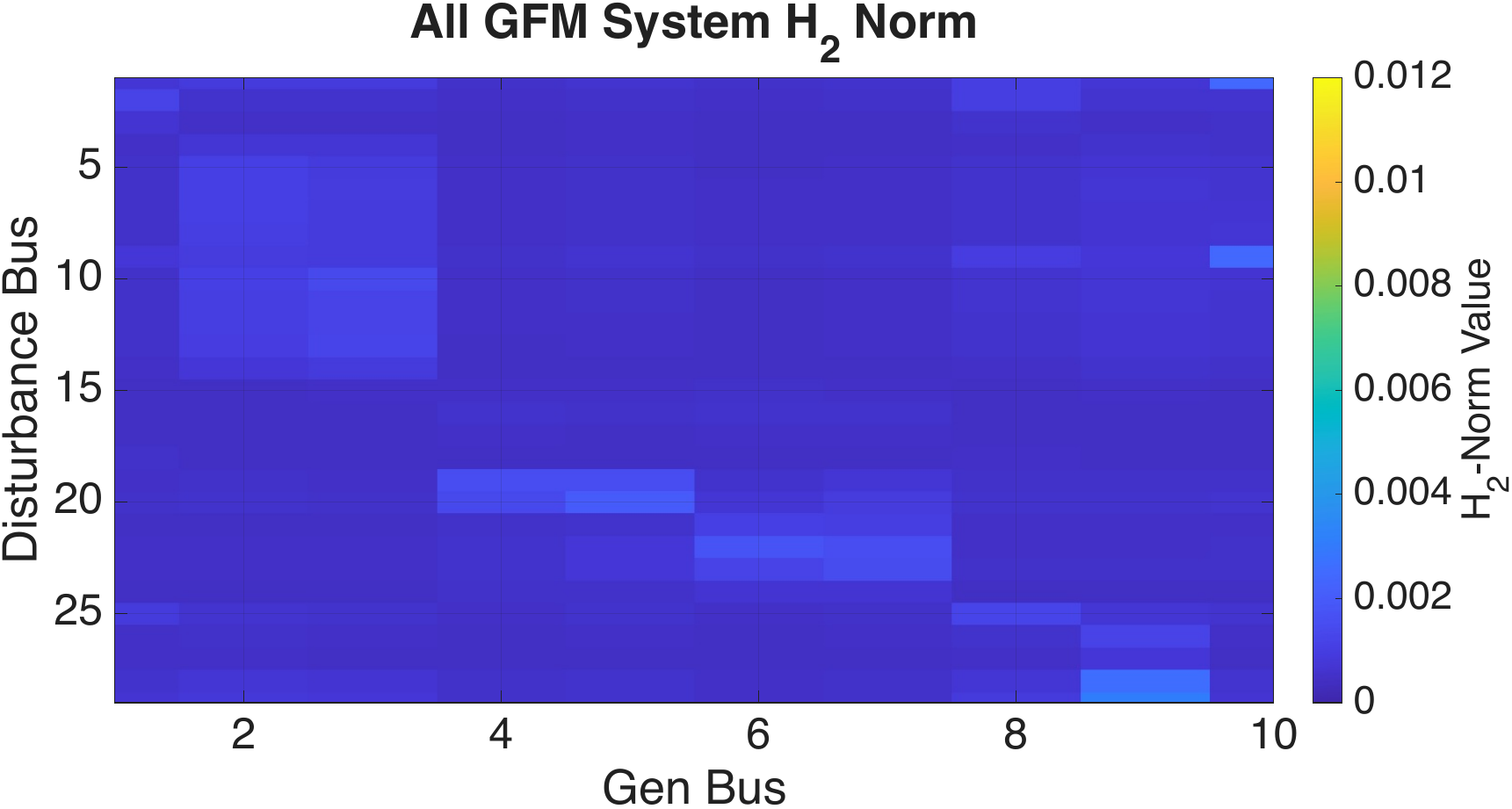}}
    \caption{IEEE 39-Bus  Map of all-GFM system with slower time constant}
    \label{fig:h2_norm_map_slow} \vspace{-2mm}
\end{figure}

To further quantify the relationship between the disturbance location and the response of a generator, the $\mathcal{H}_2$-norm from each load bus input to each generator frequency output is calculated. Each input-output  pair is plotted on a 2-D surface plot where each entry in the map represents how reactive a generator (the $x$-coordinate in the map) is to a disturbance at a load bus (the $y$-coordinate in the map). The calculated $\mathcal{H}_2$-norms for both the all-SG and all-GFM cases of the 39-bus system are shown in Figure \ref{fig:h2_norm_map}. 
For the all-GFM system, the $\mathcal{H}_2$-norm from load buses and generators closer together are significantly greater than $\mathcal{H}_2$-norm from load buses to generators farther away. This suggests that the all-GFM system will produce a more localized disturbance response. This trend is not seen in the heatmap of the $\mathcal{H}_2$-norms for all-SG system (Figure~\ref{fig:h2_norm_map}-b) where the entries are roughly the same, meaning that each generator in the system will respond similarly to a load step regardless of location in the network. To further illustrate the effect of low inertia on localization, the time constant of the GFMs is increased by a factor of 10, creating an effective inertia closer to that of SGs. It should be noted, however, that although the effective inertia is similar to that of a SG, the GFM step response shape -- dictated by the model in the lower part of Figure \ref{fig:SG_block} -- differs from that of a SG, which includes an additional feedback loop. The calculated $\mathcal{H}_2$-norms are depicted by the heatmap in Figure \ref{fig:h2_norm_map_slow}.
 The recalculated $\mathcal{H}_2$-norm of the load bus to generator mappings are visualized by a heatmap lacking the peaks seen in the low inertia GFM case, creating a similar structure to the all-SG case.  This indicates that the localization of load step responses in all-GFM systems is driven by the speed of the GFM response caused by low effective inertia.
 
\begin{figure}[h!]
    \centering
    \subfigure[] {\includegraphics[width=0.45\linewidth]{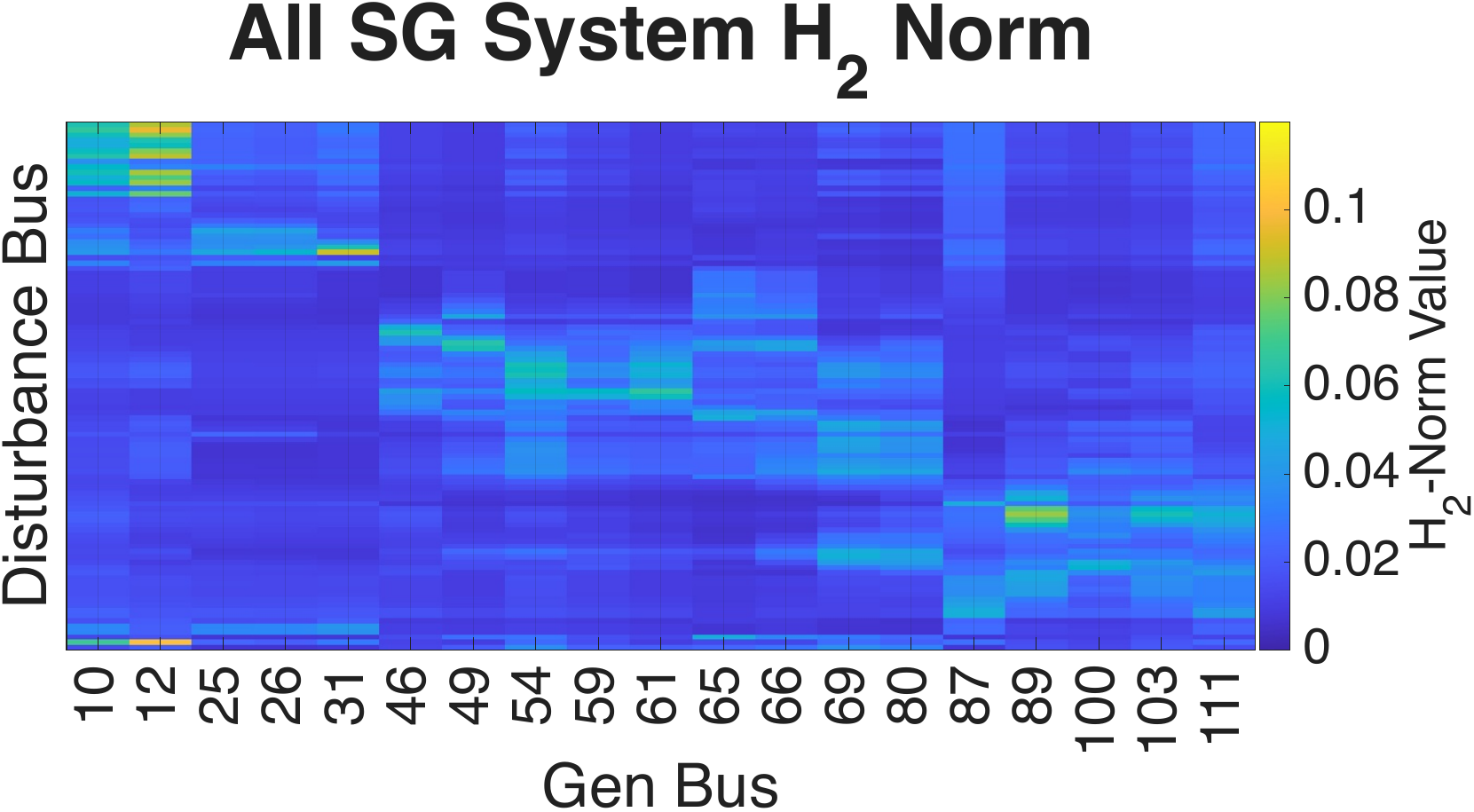}}
     \subfigure[] {\includegraphics[width=0.45\linewidth]{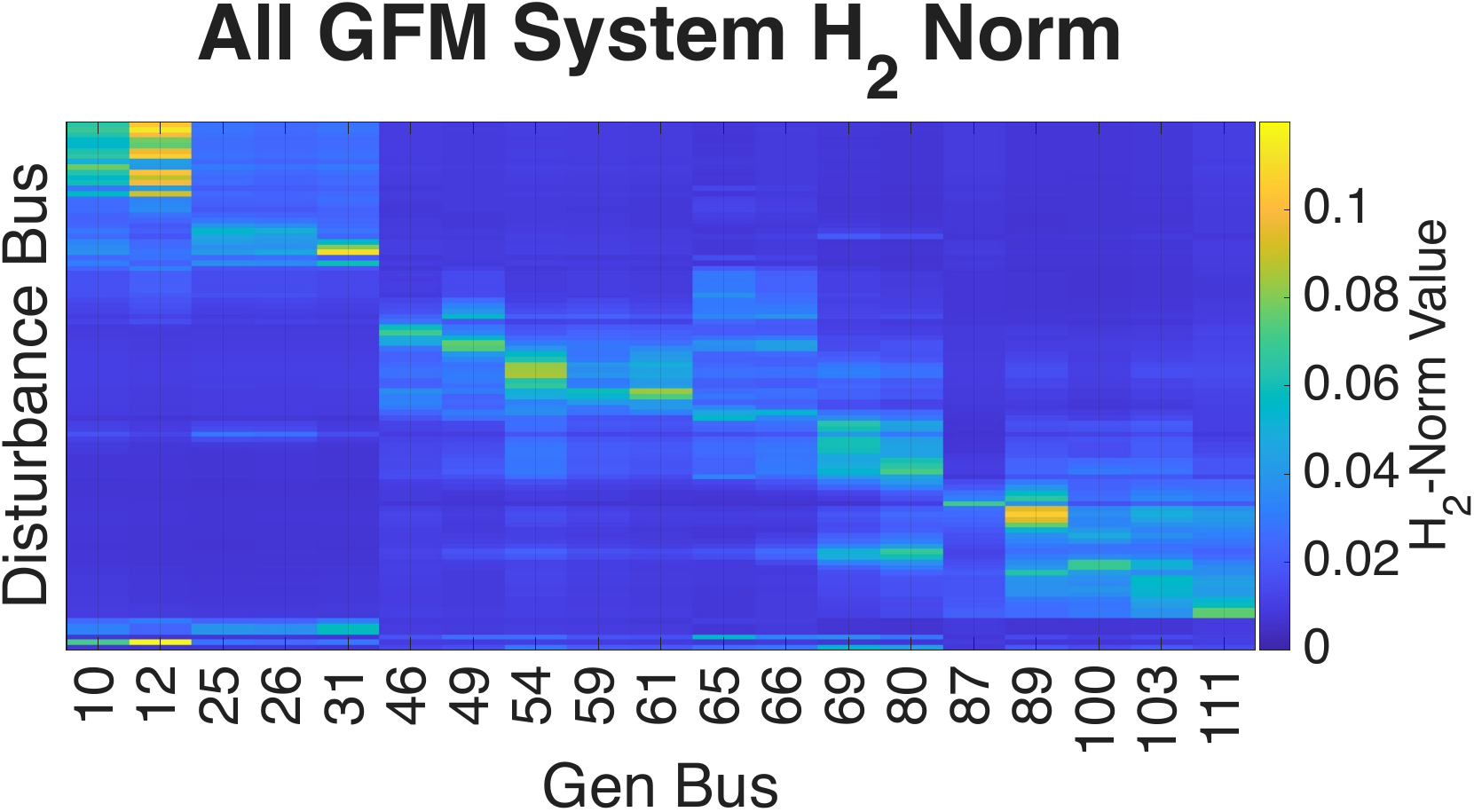}}
    \caption{IEEE 118-Bus  Map of (a) All-SG System and (b) All-GFM System}
    \label{fig:118_bus_h2_norm} \vspace{-2mm}
\end{figure}

 To evaluate localization behavior with larger power networks, the $\mathcal{H}_2$-norm is calculated for all-SG and all-GFM instantiations of the IEEE 118-bus system. The base 118-bus system is modified to remove any synchronous condensers and all other parameters were left unchanged \cite{IEEE118BusSystem}.  The resultant $\mathcal{H}_2$-norm heatmaps are shown in Figure \ref{fig:118_bus_h2_norm}.

Once again, the all-GFM  heatmap displays higher localized peaks compared to the all-SG case, indicating that the expected localization of responses scales to larger systems. When the all-GFM 118-bus case is rerun with increased effective inertia, the peaks in the $\mathcal{H}_2$-norm heatmap are removed, leading to a heatmap that resembles that of the all-SG case. This further supports the conjecture that the lower effective inertia of GFMs drives the localization of responses, even in larger power systems.
\subsection{Controllability Analysis}
\begin{figure}[h!]
    \centering
    \includegraphics[width=0.9\linewidth]{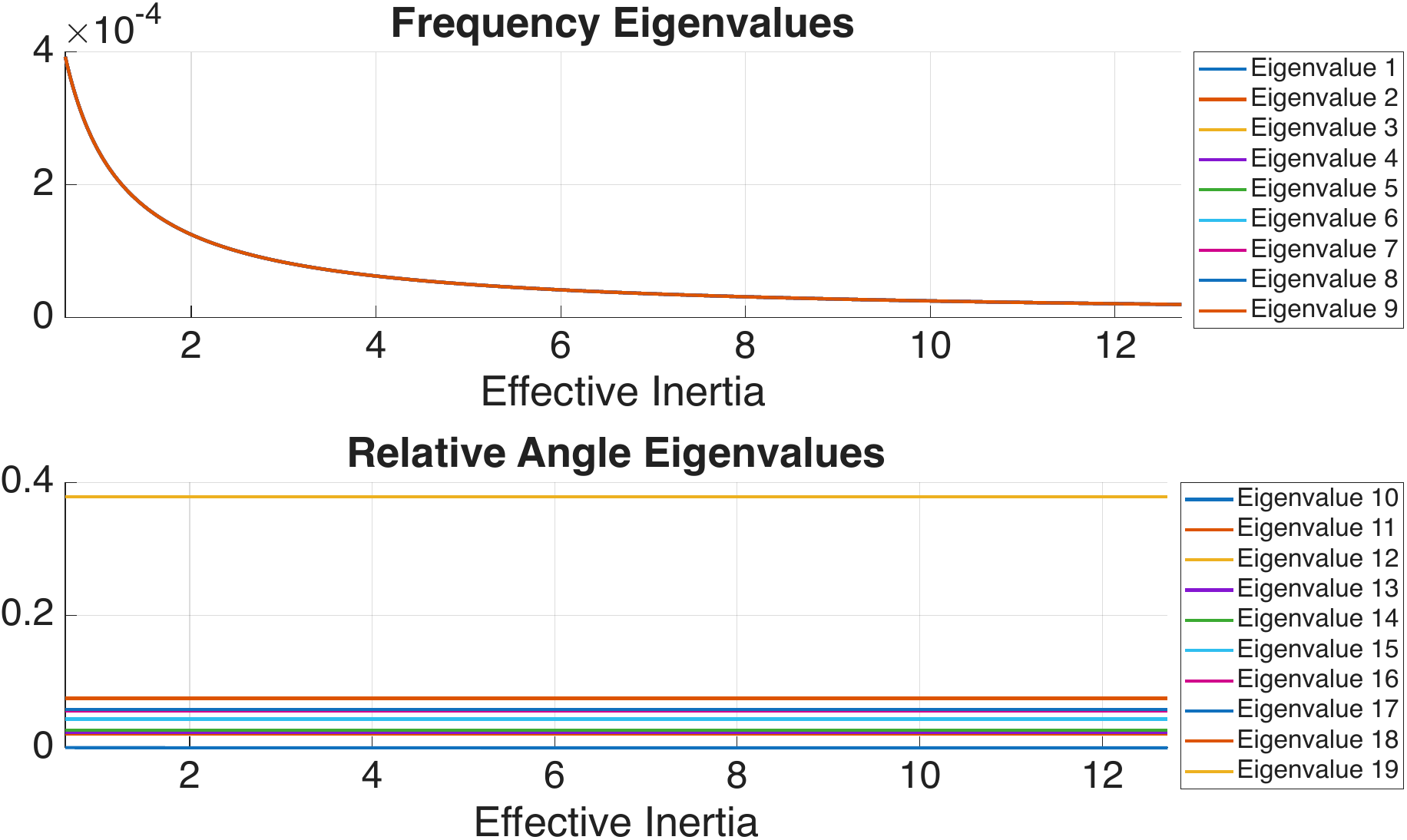}
    \caption{Controllability Gramian eigenvalues vs. effective inertia for effective inertia values ranging from $0.636$ to $12.73$.}
    \label{fig:39bus_eig_trend}
\end{figure}

\indent To understand how the conclusions of Theorem 1 change in a more realistic network topology, the controllability Gramian of the all-GFM 39-bus system model and its eigenvalues were calculated for effective inertia values from $0.636$ to $12.73$ under the assumption that all GFMs in the system are identical. The trend of the gramian eigenvalues as inertia is increased is shown in Figure \ref{fig:39bus_eig_trend}.  Even \textcolor{black}{when line susceptances are not uniform}, the controllability Gramian maintains the block diagonal structure used to derive the eigenvalues in our proof of Theorem 1.  Therefore, the eigenvalues can be broken up into those related to controlling relative GFM phase and those related to controlling GFM frequency.
Figure \ref{fig:39bus_eig_trend} shows that the eigenvalues associated with controlling frequencies are identical and scale inversely with GFM effective inertia, aligning with the conclusion drawn using a simplified network topology.  
\textcolor{black}{Although the controllability Gramian eigenvalues corresponding to relative angles do not exactly match the analytical expressions given in Theorem 1 due to the more complex network topology, they are still observed to be invariant under changes to GFM inertia as shown in Figure~\ref{fig:39bus_eig_trend}.}
This result supports the main takeaway of Section \ref{sec:analytical_controllability} -- that the lower inertia of GFMs make the frequencies of the GFMs easier to control.  These findings will influence future work on WAC system design as actuation on GFMs will be more effective at controlling system frequencies due to the lower inertia properties.
\section{Conclusion}\label{sec:conc}
This paper demonstrates that the limited effective inertia of droop GFMs leads to greater disturbance response localization in a system while also increasing the overall controllability of the inverter frequencies. The greater disturbance localization in GFM-dominated systems implies that measurements used for feedback in a WAC system will need to be close to generator nodes to capture GFM frequency response, while the increase in controllability in GFM-dominated systems implies that WAC systems will be more effective when actuating at GFMs. These takeaways motivate future work on developing a design method for the optimal implementation of WAC systems in modern power systems. \textcolor{black}{Next steps in this line of work include extension of the analysis of the localization and controllability properties to networks that are made up of both SGs and GFMs.}
\bibliographystyle{IEEEtran}
\bibliography{disturbance_localization_bibliography}

\appendix

\section{Appendix A Title}
\subsection{Proof of Proposition~\ref{prop:H2}}
We use the fact that the $\mathcal{H}_2$-norm of a system $G$ with state space realization $G = \left[ \begin{array}{c|c} A & B \\ \hline C & 0 \end{array} \right]$ is given by ${\rm Tr}(C W C^{\top})$ where $W$ is the controllability Gramian of $(A,B). $
State space realizations of $T_{\rm d}$ and $T_{\rm od}$ are given by $\left[ \begin{array}{c|c} A_T & B_T \\ \hline C_{\rm d} & 0 \end{array} \right]$ and $\left[ \begin{array}{c|c} A_T & B_T \\ \hline C_{\rm od} & 0 \end{array} \right]$, respectively where $
    A_T = {\small{\lba{ccc} 0 & 1 & 0 \\ 0 & 0 & 1 \\ \tfrac{-3 D b \omega_0 \alpha}{M^2}  & - \left( \tfrac{3b\alpha \omega_0}{M} + \tfrac{D^2}{M^2}\right) & \tfrac{-2D}{M}\ear}},  B_T = {\small{\lba{c} 0 \\ 0 \\ 1 \ear}}$,  $C_{\rm od} = \lba{ccc} \tfrac{b \alpha^2 \omega_0}{M^2} & 0 & 0 \ear, 
$ and $C_{\rm d} = \tfrac{\alpha}{M} \lba{ccc} \tfrac{b \alpha \omega_0}{M} & \frac{D}{M} & 1 \ear$. The remaining straightforward computations are omitted. 
\subsection{Proof of Theorem~\ref{main_thm}}
\color{black}
This result follows from the computation of the system's controllability Gramian, $W_c$. 
{Controllability of the system $(A,B)$ ensures that $W_c$ is the unique positive definite solution to \eqref{eq:ctrb_gram}}. 
We write $W_c$ as $W_c = {\small{\begin{bmatrix} W_{11} & W_{12} \\ W_{12}^{\top} & W_{22} \end{bmatrix}}}$ and block partition the other matrices in \eqref{eq:ctrb_gram} to write: 
{\small{\begin{equation} \label{eq:Wc_eqn}
 \begin{bmatrix}\mathbb{0} & A_{12} \\A_{21} & A_{22}  \end{bmatrix} \begin{bmatrix} W_{11} & W_{12} \\ W_{12}^{\top} & W_{22} \end{bmatrix} + \begin{bmatrix} W_{11} & W_{12} \\ W_{12}^{\top} & W_{22} \end{bmatrix} \begin{bmatrix} \mathbb{0} & A_{21}^{\top} \\ A_{12}^{\top} & A_{22}^{\top}\end{bmatrix} = \begin{bmatrix}
     \mathbb{0} & \mathbb{0} \\ \mathbb{0} & \tfrac{1}{M^2}I
 \end{bmatrix}
\end{equation}}}
where $A_{12} = \omega_0\begin{bmatrix} I_{n-1} & \mathbb{1}_{n-1}\end{bmatrix},$ $A_{21} = \frac{-\alpha }{M_{eff}}\boldsymbol{\tilde{B_{r}}}$, and $A_{22} = \frac{-D_{eff}}{M_{eff}}I_n$.
Then direct substitution of 
\begin{equation} \label{eq:Wc}
    \begin{bmatrix} W_{11} & W_{12} \\ W_{12}^{\top} & W_{22} \end{bmatrix}  = \begin{bmatrix}
    \begin{matrix}
        \frac{\omega_0}{2nbD}(\mathbb{1}_{n-1 \times n-1}+I_{n-1})
    \end{matrix} & \mathbb{0}_{n-1\times n}\\
    \mathbb{0}_{n\times n-1} & \begin{matrix}
        \frac{1}{2MD}I_n
    \end{matrix}
    \end{bmatrix} \succ 0
\end{equation}
into \eqref{eq:Wc_eqn} confirms that this is the (unique) solution and thus defines the system's controllability Gramian. 

Since $W_C$ is block diagonal, its eigenvalues are simply those of its two diagonal blocks: the upper-left block, associated with the first $n\!-\!1$ states (GFM relative phase angles), and the lower-right block, associated with the remaining $n$ states (GFM frequencies). The upper-left block has one eigenvalue equal to $\frac{\omega_0}{2bD}$ and $n-2$ eigenvalues equal to $\frac{\omega_0}{2nbD}$. The lower-right block is diagonal, so its eigenvalues are its diagonal entries, yielding $n$ eigenvalues of $\frac{1}{2MD}$. \hfill $\square$
\end{document}